\documentclass[10pt,twocolumn]{ICCAS2023}
 
\usepackage{diagbox}
\usepackage{cite}

\begin{document}

\title{Empirical Modeling of Variance in Medium Frequency R-Mode Time-of-Arrival Measurements}

\author{Jaewon Yu${}^{1}$ and Pyo-Woong Son${}^{2,3*}$ }

\affils{ ${}^{1}$School of Integrated Technology, Yonsei University, \\
Incheon, 21983, Korea (jaewon.yu@yonsei.ac.kr) \\
${}^{2}$Korea Research Institute of Ships and Ocean Engineering, \\
Daejeon, 34103, Korea (pwson@kriso.re.kr) \\
${}^{3}$Ship and Ocean Engineering Major, University of Science and Technology, \\
Daejeon, 34113, Korea \\
{\small${}^{*}$ Corresponding author}}

%\thanks{ \noindent
%   This paper is supported by my funding agencies.
%  }

\abstract{
The R-Mode system, an advanced terrestrial integrated navigation system, is designed to address the vulnerabilities of global navigation satellite systems (GNSS) and explore the potential of a complementary navigation system. 
This study aims to enhance the accuracy of performance simulation for the medium frequency (MF) R-Mode system by modeling the variance of time-of-arrival (TOA) measurements based on actual data. 
Drawing inspiration from the method used to calculate the standard deviation of time-of-reception (TOR) measurements in Loran, we adapted and applied this approach to the MF R-Mode system. 
Data were collected from transmitters in Palmi and Chungju, South Korea, and the parameters for modeling the variance of TOA were estimated.
}


\maketitle

%-----------------------------------------------------------------------

\section{Introduction}

The Global Navigation Satellite Systems (GNSS) \cite{Kim14:Comprehensive, DeLorenzo10:WAAS/L5, Chen11:Real, Lee23:Seamless, Kim23:Machine, Kim23:Low}, which derive user positioning from signals received from satellites, such as the United States' GPS \cite{Enge11:Global} and Europe's Galileo, are vulnerable to radio frequency interference (RFI) \cite{Park21:Single, Park18:Dual, Kim19:Mitigation, Schmidt20:A, Park17:Adaptive, Jeong20:RSS} and ionospheric anomalies \cite{Jiao15:Comparison, Lee17:Monitoring, Seo11:Availability, Lee22:Optimal, Sun21:Markov, Sun20:Performance, Ahmed17:Statistical}, primarily due to the long distance from satellites to the Earth's surface, resulting in weakened signal strength. 
Instances of GNSS RFI have been reported globally, with notable cases in South Korea where North Korea has intentionally jammed GPS signals \cite{Kim22:First, Rhee21:Enhanced, Son18:Novel, Son19:Universal, Son20:eLoran}.

In response to these challenges, South Korea is developing an alternative navigation system called ``R-Mode'' for maritime users to use in the event of GNSS failure. 
The R-Mode system \cite{Son22:Analysis, Jeong22:Preliminary, Son23:Skywave}, a terrestrial integrated navigation system, processes eLoran \cite{Son23:Demonstration, Son18:Preliminary, Williams13:UK, Pelgrum06:New, Li20:Research, Kim20:Development, Park20:Effect, Hwang18:TDOA, Son19:Preliminary} signals and other signals to calculate positions when GNSS signals are unavailable. 
The recent focus of R-Mode research is on methods utilizing medium frequency (MF) or very high frequency (VHF) signals for position calculation \cite{Johnson14:Feasibility1, Johnson14:Feasibility2, Johnson14:Feasibility3}.

To support the deployment of the MF R-Mode system in South Korea, the development of a simulation tool capable of predicting the system's navigation performance is necessary. 
The MF R-Mode simulation tool should be able to estimate signal strength and noise at a given location to calculate the signal-to-noise ratio (SNR). 
The variance of time-of-arrival (TOA) measurements is essential for estimating positioning accuracy. 
While a method to simulate the strength of the MF R-Mode signals has been proposed \cite{Yu22:Simulation}, a mathematical formula with appropriate parameters that relates the variance of TOA measurements to SNR for the MF R-Mode is not yet determined. 
In this study, we utilized the variance formula of the eLoran system but estimated parameters that are suitable for the Korean MF R-Mode testbed system based on actual MF R-Mode signal measurements.

\section{Methodology}

\subsection{Variance Formula for eLoran TOR Measurements}

In the eLoran system, the standard deviation of the bias-removed time-of-reception (TOR) measurements from a transmitter, denoted as $\sigma_{i}$, is a function of the transmitter's jitter ($J_{i}$) and the signal-to-noise ratio ($S\!N\!R_{i}$) of the received signals, as illustrated by the following equation \cite{Lo08:Loran, Rhee21:Enhanced}:
\begin{equation}
\sigma_{i}^{2} = J_{i}^{2} + \frac{337.5^{2}}{N_{\mathrm{pulses}} \cdot S\!N\!R_{i}}
\end{equation}
Here, $N_{\mathrm{pulses}}$ represents the number of accumulated pulses of the Loran signal, determined by the group repetition interval (GRI) of the respective Loran chain.
The number 337.5 was derived from a benchmark measurement \cite{Lo08:Loran}.

Various factors can influence the transmitter jitter, such as thermal noise, bandwidth limitation, improper impedance termination, asymmetries in rise and fall times, and cross-coupling.
It is essential to note that each transmitter has a different jitter value.
Thus, the actual jitter should be estimated based on measurements \cite{Rhee21:Enhanced}.

\subsection{Estimation of Parameters for the Variance Formula of MF R-Mode TOA Measurements}

We drew inspiration from the conventional formula for calculating the standard deviation ($\sigma_i$) of Loran's TOR measurements and applied it to MF R-Mode.
The modified formula for MF R-Mode, which calculates the variance of MF R-Mode TOA measurements, is as follows:
\begin{equation}
\label{eq:model}
\sigma_i^2 = J_i^2 + \frac{C^2}{S\!N\!R_i}
\end{equation}
where $J_i$ represents the jitter of transmitter $i$, $C$ is a constant, and $S\!N\!R_i$ is the signal-to-noise ratio of the received signals from transmitter $i$. 

We estimated $J_i$ and $C$ simultaneously based on actual MF R-Mode measurement data.
The estimation method considered the point at which the residual sum of squares (RSS) between the actual measurements and the model curve was at its minimum.
Minimizing RSS is a typical method in parameter estimation problems.

\subsection{Data Acquisition and Processing for Parameter Estimation}

In the data acquisition process, the ``MFR-1a Medium Frequency R-Mode Receiver'' by Serco was utilized to collect MF R-Mode signals from the transmitters in Palmi and Chungju, South Korea.
This Serco receiver was also used in \cite{Johnson20:R-Mode}.
The Serco receiver provides raw phase measurements and SNRs of the received MF signals, which are pivotal in subsequent analyses.

The $S\!N\!R_i$ for transmitter $i$ in (\ref{eq:model}) is measured by the receiver, but $\sigma_i^2$ in (\ref{eq:model}) is not directly measured.
Thus, it is necessary to derive $\sigma_i^2$ from the raw phase measurements $\phi_{{\mathrm raw}, i}$.
Since the measured $\phi_{{\mathrm raw}, i}$ values fall within the range of 0 and $2 \pi$, they can exhibit sudden discontinuities due to measurement noise if $\phi_{{\mathrm raw}, i}$ is close to 0 or $2 \pi$.
In cases where there is a phase discontinuity larger than $\pi$ between two adjacent epochs, the phase value is adjusted by $2 \pi$ to ensure the continuity of phase values.
This adjusted phase for transmitter $i$ is denoted as $\phi_{{\mathrm cont}, i}$.

The relationship between $T\!O\!A_i$ for transmitter $i$ and $\phi_{{\mathrm cont}, i}$ is expressed as follows:
\begin{equation}
{T\!O\!A_i} = \left(\frac{\phi_{{\mathrm cont}, i}}{2 \pi}\right)\lambda + n \lambda
\end{equation}
where $n$ represents the number of complete cycles of the MF continuous wave (CW) signal from the transmitter to the receiver, $\lambda$ is the wavelength of the CW signal, and $T\!O\!A_i$ denotes the time-of-arrival or range for transmitter $i$.

Given that the $n \lambda$ distance between transmitter $i$ and the stationary receiver remains constant, the variance of TOA, denoted as $\sigma_i^2$ in (\ref{eq:model}), can be calculated as follows:
\begin{equation}
\sigma_i^2
= {\mathrm Var} \left( {T\!O\!A_i} \right)
= \left(\frac{\lambda}{2 \pi}\right)^2 \cdot {\mathrm Var} \left( \phi_{{\mathrm cont}, i} \right)
\end{equation}

Now, the $S\!N\!R_i$ and $\sigma_i^2$ for transmitter $i$ in (\ref{eq:model}) are obtained based on the receiver measurements.
Subsequently, the values of $J_i^2$ and $C^2$ can be estimated by fitting the model in (\ref{eq:model}) to the measurements.

\section{Results}

\begin{figure}
\centering
   \includegraphics[width=1.0\linewidth]{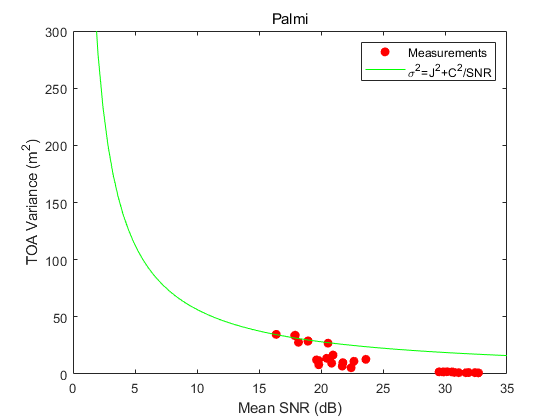}
\caption{Estimation of model parameters using measurements for the Palmi transmitter.}
\label{fig:Palmi}
\end{figure}

\begin{figure}
\centering
   \includegraphics[width=1.0\linewidth]{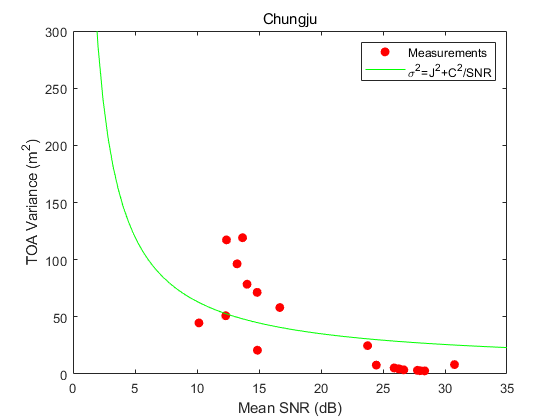}
\caption{Estimation of model parameters using measurements for the Chungju transmitter.}
\label{fig:Chungju}
\end{figure}

In Figs. \ref{fig:Palmi} and \ref{fig:Chungju}, the red dots represent actual measurement values, with $S\!N\!R_i$ plotted along the \textit{x}-axis and the variance of TOA (i.e., $\sigma_i^2$) along the \textit{y}-axis.
The green curve corresponds to the model presented in (\ref{eq:model}), with the calculated values of $J_i$ and $C$ that yield the best fit by minimizing the RSS between the model and the measurements.
Through this process, we determined that $C = 23.75$, $J_{\mathrm{Palmi}} = 0.00$, and $J_{\mathrm{Chungju}} = 2.65$ achieved the minimum RSS.
Using these determined values of $J_i$ and $C$, $\sigma_i^2$ can be predicted for a given $S\!N\!R_i$ using the relation presented in (\ref{eq:model}).

\section{Conclusion}

In this study, we estimated the parameters ($J_i$ and $C$) of the variance model for MF R-Mode TOA based on actual measurements of $S\!N\!R_i$ and $\phi_{{\mathrm raw}, i}$. 
The obtained variance model predicts realistic $\sigma_i^2$, which is necessary for positioning accuracy simulation. 
Therefore, this study contributes to improving the performance simulation capability for the MF R-Mode system under development in South Korea.

\section*{ACKNOWLEDGEMENT}

This research was conducted as a part of the project titled ``Development of integrated R-Mode navigation system [PMS4440]'' funded by the Ministry of Oceans and Fisheries, Republic of Korea (20200450).
This research was also supported by the Future Space Navigation and Satellite Research Center through the National Research Foundation of Korea (NRF) funded by the Ministry of Science and ICT, Republic of Korea (2022M1A3C2074404).

\bibliographystyle{IEEEtran}
\bibliography{mybibfile, IUS_publications}

\end{document}